\documentclass[10pt]{article}
\makeatletter
\usepackage{setspace}
\usepackage{rotating}
\usepackage{multirow}
\usepackage{mathtools}
\usepackage{ragged2e}
\usepackage{tikz}
\usepackage{longtable}
\usetikzlibrary{shapes.geometric, arrows}

\tikzstyle{startstop} = [rectangle, rounded corners, 
text centered, 
draw=black]

\tikzstyle{io} = [trapezium, 
trapezium stretches=true, 
trapezium left angle=70, 
trapezium right angle=110,text centered, 
draw=black,text width=3cm,]

\tikzstyle{process} = [rectangle,  
draw=black,text width=1.5cm, text centered]

\tikzstyle{decision} = [diamond, 
draw=black,text width=1.5cm, text centered]
\tikzstyle{arrow} = [thick,->,>=stealth]
\usepackage{epsfig}
\usepackage{rotating}
\usepackage{amssymb}
\usepackage{xcolor}
\usepackage[T1]{fontenc}
\usepackage{lmodern}
\usepackage[utf8]{inputenc}
\newenvironment{proof}{\paragraph{Proof:}}{\hfill(Proved)}
\usepackage{booktabs}
\usepackage{multirow}
\usepackage{enumerate}
\usepackage{caption}
\usepackage{subcaption}
\usepackage{lscape,float}
\usepackage{graphicx}
\usepackage{algorithm} 
\usepackage{algpseudocode}
\usepackage[utf8]{inputenc}
\usepackage{xcolor}
\usepackage{soul}
\usepackage[toc]{appendix}

\usepackage{natbib,hyperref,doi}
\makeatletter
\renewcommand\@biblabel[1]{#1.}
\makeatother

\usepackage{amsmath}
\usepackage{enumitem}
\newtheorem{theorem}{Theorem}

\newtheorem{lemma}{Lemma}
\makeatletter
\renewcommand\@biblabel[1]{#1.}
\makeatother

\usepackage{float,lscape}
 \providecommand{\keywords}[1] {
  \textit{Keywords:} #1}
\title{ Bayesian Optimum Warranty Region for Right Censored Two Dimensional Dependent Data}

\author{
    Tanmay Sen$^1$\thanks{Corresponding author\\Email address: tanmay.sen@isical.ac.in} \and
    Rathin Das$^1$ \and
   Ritwik Bhattacharya$^2$ \and
}
\date{$^1$ SQC \& OR Unit, Indian Statistical Institute, Kolkata, India\\
$^2$ Department of Mathematical Sciences, University of Texas El Paso, El Paso, TX}
\usepackage{siunitx}

\begin{document}

\maketitle

\begin{abstract}

Warranty policies play a crucial role in balancing customer satisfaction and manufacturer’s cost. Traditional one-dimensional warranty frameworks, based solely on either age or usage, often fail to capture the joint effect of product life factors. This article investigates two-dimensional warranty policies by combining Free Replacement Warranty (FRW), Pro-Rata Warranty (PRW), and Combination FRW-PRW Warranty (CW) schemes across both age and usage scales. A dissatisfaction cost function is proposed alongside the economic benefit and warranty cost functions, and the expected utility framework is employed to derive optimal warranty parameters. The expectation is taken with respect to the posterior predictive distribution of product lifetime and usage data, ensuring a data-driven approach. Finally, the methodology is validated using an open-source dataset, and a new two-dimensional starter motor dataset is introduced to demonstrate the practical advantages of adopting two-dimensional warranty policies.

\end{abstract}


\section{Introduction}

Warranties serve as a key marketing tool for manufacturers and sellers in today’s competitive global market, providing assurance to buyers about the quality and reliability of their products \citep{blischke_2006}. However, determining the optimal warranty length (in one dimension, age) or the warranty region (in two dimensions, age and usage) is primarily a matter of profitability and cost management for the manufacturer. A longer warranty generally signals higher product reliability to consumers, but if actual reliability is low, an overly generous warranty can result in substantial costs. Conversely, offering a warranty period shorter than that of competitors may discourage sales. Thus, selecting an appropriate warranty policy is a critical task, typically guided by reliability assessments from life-testing experiments, often conducted under censoring schemes to reduce time and cost.  

Traditional warranty analysis is commonly based on a single time scale, usually the age of the product \citep{wu_2010, Christen_2006, sen2022determination}. However, this one-dimensional approach often fails to reflect actual product reliability, since usage varies significantly across consumers. Two products of the same age may undergo very different levels of wear depending on intensity of use. Ignoring this heterogeneity may lead to biased reliability estimates and suboptimal warranty decisions. To overcome this limitation, \emph{two-dimensional warranties} \citep{das2025determination}, which account for both age and usage (e.g., mileage, operating hours, or cycles), have been introduced as a more realistic framework.  

Another limitation of traditional approaches is their reliance on classical (frequentist) estimation methods. While these methods yield point estimates, they do not fully capture the uncertainty inherent in censored data. Bayesian approaches, by contrast, incorporate prior knowledge and provide probabilistic inference, resulting in more robust parameter estimates and improved decision-making under uncertainty. This is particularly advantageous for two-dimensional warranties, where uncertainty plays a critical role in determining optimal coverage policies.  

The lifetimes of products such as factory equipment, automobiles, traction motors, and starter motors are naturally characterized by both time and usage \citep{jung2007analysis}. These products are often sold with two-dimensional warranties, where both factors jointly determine eligibility for claims. However, analyzing such data presents challenges. In practice, warranty expiration times are not always directly observed. For example, in a 5-year/40,000-mile warranty, vehicles with high mileage may not survive the full time horizon, while vehicles with low mileage may not accumulate sufficient usage before expiration. Consequently, exact censoring times and accumulated usage are often unknown.  In practice, manufacturers usually know how many products were sold and how many returned under warranty. Failures are observed, while unreturned units can reasonably be treated as right-censored. Since explicit censoring times are rarely available, we generate censoring times from observed failure data to replicate realistic warranty scenarios where both failures and censored observations coexist.  

This study focuses on the determination of a \textit{Bayesian optimal warranty region} under a two-dimensional warranty setting. We consider a combined \emph{free replacement and pro-rata warranty (FRW–PRW) policy}, with the expected utility defined as a sum of economic benefit, warranty cost, and consumer dissatisfaction cost function. To model the positively correlated two-dimensional data, we employ the \emph{multivariate extension (ME) model} proposed by \citet{roy1998multivariate}, using the Weibull distribution to represent both the age and usage scales. The main contributions of this work are:  
\begin{itemize}
\justifying
\itemsep 1pt
    \item Proposing a two-dimensional dissatisfaction cost function that captures all possible scenarios under the combined FRW–PRW policy.  
    \item Designing the optimal warranty region within a Bayesian framework by maximizing expected utility.  
    \item Providing the first comprehensive Bayesian framework for designing two-dimensional warranty regions under right-censored field data, while incorporating distinct cost functions.  
\end{itemize}

\section{Related Works}
Determining the optimum warranty period or region is crucial, as it directly impacts the manufacturer’s cost. Without an appropriate design, the manufacturer may incur substantial losses due to the trade-off between product reliability and warranty coverage. Warranties are classified into two main types: one-dimensional, which cover only the duration of time, and two-dimensional, which consider both time and usage or other relevant factors. Many studies have focused on determining one-dimensional warranty periods for various policies, using either complete or censored data, through frequentist or Bayesian approaches. \cite{Christen_2006} is the first work to maximize the utility function by considering economic benefits, warranty costs, and dissatisfaction costs, and computed the Bayesian optimal warranty length for a pro-rata warranty policy, modeling the product lifetime with a two-parameter Weibull distribution. \cite{wu_2010} investigated a decision problem under the FRW-PRW hybrid policy, using a Bayesian approach to determine optimal warranty lengths with a Rayleigh lifetime model under a Type-II progressive censoring scheme. The work of \cite{sen2022determination} differs significantly by considering warranty and dissatisfaction costs as a non-linear function of the remaining product lifetime. 

One-dimensional warranties are based on a single factor, such as time or usage, whereas two-dimensional warranties consider both factors, providing a more accurate assessment of product wear and allowing precise calculation of the warranty period and cost. \citet{manna2006optimal} investigated the optimal determination of the warranty region for a two-dimensional FRW policy under a specified budget constraint on warranty cost. While \citet{manna2007use} proposed a two-dimensional failure probability model indexed by age and usage to study automobile warranty problems, emphasizing how use-rate affects life and illustrating warranty cost estimation with numerical examples, highlighting challenges with censored data. Further, \citet{manna2008note} examined warranty cost estimation for rectangular 2D policies, revealing discrepancies between 1D and 2D approaches and limitations of existing formulae. The authors in \cite{jung2007analysis}, proposes a method for parameter estimation of a bivariate distribution using that accounts for the positive dependence between age and usage in two-dimensional warranty claims. \citet{dai2017field} developed a field reliability model based on two-dimensional warranty data with censoring times by treating usage rate as a random variable and estimating parameters through an accelerated failure time (AFT) model combined with a stochastic EM algorithm. Recently, \cite{das2025determination} proposed linear economic benefit and warranty cost functions for two-dimensional warranties, considering all scenarios under complete sample data. They developed a method to determine the optimal two-dimensional warranty region for age and mileage under FRW-PRW policies, using a {bivariate Gumbel–Weibull model}.

Although most existing works focus on estimating reliability model parameters from two-dimensional field warranty data, with or without censoring, the integration of parameter estimation with the determination of optimal two-dimensional warranties under a Bayesian framework remains largely unexplored.

The remainder of this article is organized as follows. Section~\ref{model} models two-dimensional, positively correlated age and usage data using the ME model with Weibull-distributed marginals and details the prior and posterior distributions. Section~\ref{policy} presents the two-dimensional warranty policy. Section \ref{cost} introduces three cost functions: the economic benefit function, the warranty cost function of \citet{das2025determination}, and an additional indirect cost—the dissatisfaction cost. Section \ref{optimal} outlines the procedure for determining the optimal warranty region, while Section \ref{numerical} illustrates the proposed model using two real-life datasets. Finally, Section \ref{conclusion} concludes with a summary of findings and potential directions for future research.
\section{Lifetime Models and Posterior Distribution}\label{model}
Suppose $n$ identical items are placed on a life testing experiment. For each item, we observe two lifetime-related quantities: the \emph{age} $(T)$ and the \emph{usage} $(U)$ of the unit. Empirical evidence suggests that these two characteristics are typically dependent in practice. In this section, we introduce a flexible lifetime model that captures the joint behavior of $(T,U)$, followed by the corresponding likelihood function, prior and posterior distributions, and the Fisher information matrix.  

\subsection{Lifetime Model}\label{life}

Let $T$ and $U$ each follow Weibull distributions with cumulative hazard rates (CHRs)
\[
H_{T}(t) = \left(\frac{t}{\eta_T}\right)^{\lambda_T},
\qquad
H_{U}(u) = \left(\frac{u}{\eta_U}\right)^{\lambda_U},
\]
where $\lambda_T,\lambda_U > 0$ are the shape parameters and $\eta_T,\eta_U > 0$ are the scale parameters of the marginal distributions. To model the dependence between $(T,U)$, we employ the \emph{multivariate extension (ME)} method of \citet{roy1998multivariate}. Under this construction, the joint reliability function is
\[
R(t,u\mid\boldsymbol{\psi})
= \Pr(T \geq t, U \geq u \mid \boldsymbol{\psi})
= \exp\!\left\{-\left[\left(\frac{t}{\eta_T}\right)^{\lambda_T/\theta}
+ \left(\frac{u}{\eta_U}\right)^{\lambda_U/\theta}\right]^{\theta}\right\}, 
\quad t,u \geq 0,
\]
where $\boldsymbol{\psi}=(\eta_T,\lambda_T,\eta_U,\lambda_U,\theta)$ denotes the vector of parameters and $0<\theta\leq 1$ is the dependence parameter. This specification corresponds to the Gumbel copula with dependence parameter $1/\theta$. When $\theta=1$, the joint reliability factorizes into the product of the marginals, implying independence between $T$ and $U$. As $\theta\to 0$, the dependence becomes stronger, representing an increasingly positive association between the two lifetimes. Differentiating the reliability function twice yields the joint probability density function (PDF):
\begin{align*}
f(t,u\mid\boldsymbol{\psi})
&= \frac{\lambda_T\lambda_U}{\eta_T^{\lambda_T/\theta}\eta_U^{\lambda_U/\theta}}
\, t^{\lambda_T/\theta - 1}\, u^{\lambda_U/\theta - 1} \,
\left[\left(\frac{t}{\eta_T}\right)^{\lambda_T/\theta} 
+ \left(\frac{u}{\eta_U}\right)^{\lambda_U/\theta}\right]^{\theta-2} \\
&\quad \times \left\{\left[\left(\frac{t}{\eta_T}\right)^{\lambda_T/\theta} 
+ \left(\frac{u}{\eta_U}\right)^{\lambda_U/\theta}\right]^{\theta} 
- \frac{\theta-1}{\theta}\right\}
\exp\!\left\{-\left[\left(\frac{t}{\eta_T}\right)^{\lambda_T/\theta} 
+ \left(\frac{u}{\eta_U}\right)^{\lambda_U/\theta}\right]^{\theta}\right\}, \quad t,u>0.
\end{align*}
The joint cumulative distribution function (CDF) is
\[
F(t,u\mid\boldsymbol{\psi})
= 1 - R(t,u\mid\boldsymbol{\psi})
= 1 - \exp\!\left\{-\left[\left(\frac{t}{\eta_T}\right)^{\lambda_T/\theta}
+ \left(\frac{u}{\eta_U}\right)^{\lambda_U/\theta}\right]^{\theta}\right\}.
\]

\subsection{Likelihood Function}\label{like}

Consider $n$ independent units placed on test, where the experiment is terminated at a pre-specified age $T_0$ or usage $U_0$, whichever occurs first. If a unit fails before termination, both $(t_i,u_i)$ are recorded; otherwise, the unit is right censored at $(T_0,U_0)$. Define the failure indicator
\[
\delta_i \;=\; 
\begin{cases}
1, & \text{if unit $i$ fails within }(0,T_0)\times(0,U_0),\\[4pt]
0, & \text{otherwise.}
\end{cases}
\]
The likelihood contribution of observation $(t_i,u_i,\delta_i)$ is
\[
L_i(\boldsymbol{\psi})
= \big[f(t_i,u_i\mid\boldsymbol{\psi})\big]^{\delta_i}
\big[1 - F(T_0,U_0\mid\boldsymbol{\psi})\big]^{1-\delta_i}.
\]
Accordingly, the log-likelihood contribution is
\[
\ell_i(\boldsymbol{\psi})
= \delta_i \log f(t_i,u_i\mid\boldsymbol{\psi})
+ (1-\delta_i)\log\!\big[1 - F(T_0,U_0\mid\boldsymbol{\psi})\big].
\]
Since the units are independent, the full-sample likelihood is
\[
L(\boldsymbol{\psi}\mid\mathbf{x})
= \prod_{i=1}^n L_i(\boldsymbol{\psi})
= \prod_{i=1}^n
\big[f(t_i,u_i\mid\boldsymbol{\psi})\big]^{\delta_i}
\big[1 - F(T_0,U_0\mid\boldsymbol{\psi})\big]^{1-\delta_i},
\]
with log-likelihood
\[
\ell(\boldsymbol{\psi}\mid\mathbf{x})
= \sum_{i=1}^n \ell_i(\boldsymbol{\psi}).
\]
Let us $d=\sum_{i=1}^n \delta_i$ denote the total number of observed failures. Then
\begin{align}\label{loglike}
\ell(\boldsymbol{\psi}\mid\mathbf{x})
= \sum_{i:\,\delta_i=1}\log f(t_i,u_i\mid\boldsymbol{\psi})
+ (n-d)\log\!\big[1-F(T_0,U_0\mid\boldsymbol{\psi})\big].
\end{align}
Thus, the likelihood separates naturally into contributions from failed and censored units.

\subsection{Prior and Posterior Distributions}\label{dist}

For Bayesian approach, we assign independent priors to the parameters $\boldsymbol{\psi}=(\eta_T,\lambda_T,\eta_U,\lambda_U,\theta)$. For the scale and shape parameters $\eta_T,\lambda_T,\eta_U,\lambda_U$, we assume 
$\psi_j \sim \mathrm{Gamma}(a_j,b_j)$ with PDF \[
\pi_j(\psi_j) = \frac{b_j^{a_j}}{\Gamma(a_j)}\psi_j^{a_j-1}e^{-b_j\psi_j}, \quad j=1,2,3,4.
\]
For the dependence parameter, we assume $\theta \sim \mathrm{Beta}(a_5,b_5)$ with PDF
\[
\pi_5(\theta) = \frac{\theta^{a_5-1}(1-\theta)^{b_5-1}}{B(a_5,b_5)}, \quad 0<\theta<1.
\]
Under independence, the joint prior is
\[
\pi(\boldsymbol{\psi})=\prod_{i=1}^5 \pi_i(\psi_i).
\]
The posterior distribution is then given by
\[
\pi(\boldsymbol{\psi}\mid\mathbf{x})
= \frac{L(\boldsymbol{\psi}\mid\mathbf{x})\,\pi(\boldsymbol{\psi})}
{\int_{\Psi} L(\boldsymbol{\psi}\mid\mathbf{x})\,\pi(\boldsymbol{\psi})\,d\boldsymbol{\psi}}.
\]
Posterior predictive quantities for a new observation $(t,u)$ follow directly. The predictive joint PDF is
\[
f(t,u\mid \mathbf{x})
= \int f(t,u\mid\boldsymbol{\psi})\,
\pi(\boldsymbol{\psi}\mid \mathbf{x})\,d\boldsymbol{\psi},
\]
and the corresponding predictive joint CDF is
\[
F(t,u\mid \mathbf{x})
= \int_0^t \int_0^u f(x,y\mid \mathbf{x})\,dx\,dy.
\]

\subsection{Fisher Information}\label{fisher}

We now turn to the Fisher information matrix, which will be used in the next section for applying the Metropolis–Hastings (MH) algorithm to compute the optimal warranty region by Bayesian approach. 

\begin{lemma}\label{lem:score}
The expected score vector is zero:
\[
E\!\left[\frac{\partial \ell(\boldsymbol{\psi}\mid\mathbf{x})}{\partial \psi_u}\right] = 0,
\qquad u=1,\dots,5.
\]
\end{lemma}

\begin{lemma}\label{lemma_2}
The expected Hessian is
\begin{align*}
E\!\left[\frac{\partial^2 \ell(\boldsymbol{\psi}\mid\mathbf{x})}
{\partial \psi_u \,\partial \psi_v}\right]
= -n\Bigg[ &\int_0^{T_0}\int_0^{U_0}
\partial_{\psi_u}\log f(t,u\mid\boldsymbol{\psi})\,
\partial_{\psi_v}\log f(t,u\mid\boldsymbol{\psi}) \,
f(t,u\mid\boldsymbol{\psi})\,dt\,du \\
&+ [1-F(T_0,U_0\mid\boldsymbol{\psi})]\,
\partial_{\psi_u}\log[1-F(T_0,U_0\mid\boldsymbol{\psi})]\,
\partial_{\psi_v}\log[1-F(T_0,U_0\mid\boldsymbol{\psi})]\Bigg].
\end{align*}
\end{lemma}
Combining Lemmas \ref{lem:score} and \ref{lemma_2}, we obtain the following result.

\begin{theorem}\label{thm:FI}
The Fisher information matrix is
\begin{align*}
I(\boldsymbol{\psi}\mid\mathbf{x})
= n\Bigg[ &\int_0^{T_0}\int_0^{U_0}
\nabla_{\boldsymbol{\psi}}\log f(t,u\mid\boldsymbol{\psi})\,
\nabla_{\boldsymbol{\psi}}\log f(t,u\mid\boldsymbol{\psi})^{\!\top}
f(t,u\mid\boldsymbol{\psi})\,dt\,du \\
&+ [1-F(T_0,U_0\mid\boldsymbol{\psi})]\,
\nabla_{\boldsymbol{\psi}}\log[1-F(T_0,U_0\mid\boldsymbol{\psi})]\,
\nabla_{\boldsymbol{\psi}}\log[1-F(T_0,U_0\mid\boldsymbol{\psi})]^{\!\top}
\Bigg].
\end{align*}
\end{theorem}



\section{Warranty Policy}\label{policy}

Warranty policies are contractual agreements between manufacturers and consumers that provide compensation in the event of product failure within a specified period. The most widely used policies are the \emph{free replacement warranty} (FRW) \citep{murthy1992product}, the \emph{pro-rata warranty} (PRW)\cite{menke1969determination}, and their combination (FRW--PRW) \citep{blischke_2006,thomas1983optimum} as CW. A defining feature of any warranty policy is that if a product fails during the warranty period, the consumer receives compensation, either in full or on a prorated basis depending on the product’s age and usage at failure.  

Under a FRW policy, a non-repairable product is replaced with an identical item at no cost, while a repairable product is restored to working condition free of charge. In contrast, under a PRW policy, the manufacturer provides compensation that is proportional to the remaining life of the product, such that older or more heavily used items receive smaller reimbursements. In practice, hybrid schemes that combine FRW and PRW features are often implemented to balance consumer protection and warranty costs. In this paper, we focus on the \emph{combined FRW-PRW} (CW) policy in two dimensions, and our objective is to determine the optimal warranty region under right-censored field data observed on both the age and usage scales.  

\subsection*{Structure of the (Combined FRW--PRW) CW Policy}

\begin{figure}[h!]
\centering
\begin{tikzpicture}[scale=1]
\draw[->] (0,0) -- (6,0) node[right] {Age $t$};
\draw[->] (0,0) -- (0,6) node[above] {Usage $u$};

\draw[dashed] (2,0) -- (2,6) node[above] {$t_{w_1}$};
\draw[dashed] (4,0) -- (4,6) node[above] {$t_{w_2}$};
\draw[dashed] (0,2) -- (6,2) node[right] {$u_{w_1}$};
\draw[dashed] (0,4) -- (6,4) node[right] {$u_{w_2}$};

\node at (1,1) {FRW-FRW};
\node at (1,3) {PRW-FRW};
\node at (3,1) {FRW-PRW};
\node at (3,3) {PRW-PRW};
\end{tikzpicture}
\caption{Warranty regions in the two-dimensional age--usage plane under CW policy.}
\label{fig:warranty_regions}
\end{figure}
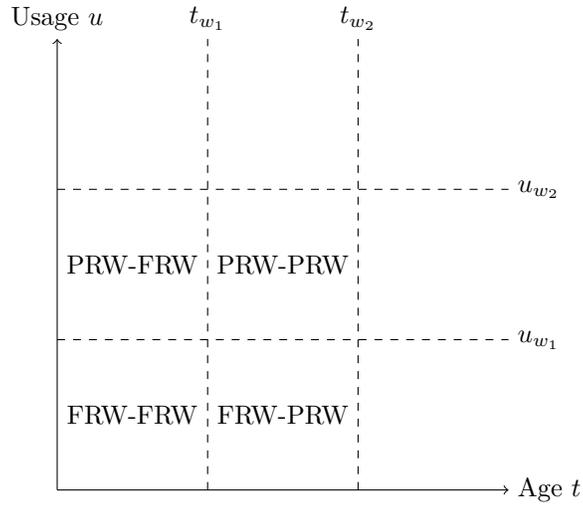

Let $t$ denote product age and $u$ denote cumulative usage. We define two warranty thresholds for each dimension: FRW thresholds $t_{w_1}$ and $u_{w_1}$, and PRW thresholds $t_{w_2}$ and $u_{w_2}$, with $t_{w_1}<t_{w_2}$ and $u_{w_1}<u_{w_2}$. We restrict attention to a rectangular warranty region in the $(t,u)$ plane \citep{wang2018two}. The CW policy partitions this region into four subregions, each associated with a distinct compensation scheme. Figure~\ref{fig:warranty_regions} illustrates this partition.  

\subsection*{Compensation Schemes by Region}
    \textit{Case 1:} FRW-FRW region
    \[
    \{(t,u): 0 < t \leq t_{w_1}, \; 0 < u \leq u_{w_1}\}.
    \]
    Failures occurring early in both age and usage are fully compensated. This region provides complete consumer protection through either free replacement (for non-repairable products) or free repair (for repairable products).  \\
    
    \noindent \textit{Case 2:} FRW-PRW region  
    \[
    \{(t,u): 0 < t \leq t_{w_1}, \; u_{w_1} < u \leq u_{w_2}\}.
    \]
    Compensation is proportional to the remaining usage of the product, such that younger and more heavily used items receive smaller reimbursements. This region captures products that are relatively new in age but heavily used.  

    \noindent \textit{Case 3:} PRW-FRW region  
    \[
    \{(t,u): t_{w_1} < t \leq t_{w_2}, \; 0 < u \leq u_{w_1}\}.
    \]
    Compensation is proportional to the remaining age of the product, such that older and lightly used items receive smaller reimbursements. This region addresses products that fail later in age but after limited use.  \\

    \noindent \textit{Case 4:} PRW-PRW region  
    \[
    \{(t,u): t_{w_1} < t \leq t_{w_2}, \; u_{w_1} < u \leq u_{w_2}\}.
    \]
    Compensation is proportional to both the remaining age and usage of the product, meaning that older and heavily used items receive the smallest reimbursements. This scheme applies to failures occurring when the product is both old and heavily used.  \\

\noindent Failures outside the region $\{t \leq t_{w_2},\, u \leq u_{w_2}\}$ are not covered by the warranty. The use of two-dimensional thresholds allows manufacturers to design more flexible and realistic warranty contracts to balance consumer satisfaction and warranty costs.  

\section{Cost Functions}\label{cost}
In this article, we consider three cost functions to determine the optimal warranty region: the economic benefit function and the warranty cost function, as proposed by \cite{das2025determination}, along with an additional indirect cost introduced in this study, namely the dissatisfaction cost. These three cost functions are used to determine the optimal warranty region by maximizing the expected utility function, where the utility function is defined by \cite{Christen_2006}, \cite{das2025determination} as  

\[
\text{Utility} = \text{Economic Benefit} - \text{Warranty Cost} - \text{Dissatisfaction Cost}.
\]
The following sections discuss these cost functions in detail.

\subsection{Economic Benefit  Function:} In a two-dimensional warranty, the economic benefit function, as defined by \cite{das2025determination}, is a monotonically increasing function that represents the manufacturer’s or seller’s gain from offering a warranty to consumers. However, excessively long warranties do not provide additional benefits; rather, an unusually large warranty compared to competitors may raise consumer doubts about the product’s reliability. Therefore, this function is bounded above. In the context of a two-dimensional warranty, \cite{das2025determination} assume that the benefit increases with the average warranty lengths across age and usage. As $t \to \infty$ and $u \to \infty$, each component benefit function converges to $1$, making their product bounded above. The economic benefit function, denoted by $EB$ is  formulated as the product of two exponential growth components, one for age and one for usage, and is bounded above by $A_1M$:  

\begin{eqnarray}\label{economic_benefit1}
    EB(t_{w_1},t_{w_2},u_{w_1},u_{w_2}) = A_1M \, \left[1 - \exp\left\{-A_2 \left(\frac{t_{w_1} + t_{w_2}}{2}\right)\right\}\right] \times \left[1 - \exp\left\{-A_3 \left(\frac{u_{w_1} + u_{w_2}}{2}\right)\right\}\right],
\end{eqnarray}




where $A_1$ represents the manufacturer’s profit per-product, and $M$ denotes the potential number of units sold under the warranty policy. The parameters $A_2$ and $A_3$ control the rate at which the benefit increases and are estimated using ratios derived from two special cases of the FRW–PRW policy. For a generic scale $x \in \{t,u\}$ with warranty length $x_w$, the ratio is defined as  
\begin{align}\label{ratio}
   h(A,x_w) = \frac{1-\exp\!\left(-\tfrac{A x_w}{2}\right)}{1-\exp\!\left(-A x_w\right)} ,
\end{align}
which is strictly increasing with $h(0^+)=0.5$ and $h(\infty)=1$.  
By selecting a reference value $q\in (0.5,1)$, the parameters $A_2$ and $A_3$ can be obtained by solving  
\[
h(A_2,t_w)=q_1, \qquad h(A_3,u_w)=q_2 ,
\]
where $q_1$ and $q_2$ denote the reference proportions for the age and usage scales, respectively.

\subsection{Warranty Cost Function:}
The warranty cost function is the direct expense borne by the manufacturer for repair, replacement, or reimbursement during the warranty period. \cite{das2025determination} considered the set of warranty policies $\mathcal{P}=\{FRW,\,PRW,\,CW\}$ for each scale (age and usage) and construct per unit warranty cost function. They have shown, a two-dimensional warranty consists of $3 \times 3 = 9$ possible scenarios, e.g., $\{FRW \times FRW,~ FRW \times PRW, ~FRW \times CW, ~PRW \times FRW, ~PRW \times PRW, ~PRW \times CW, ~CW \times FRW, ~CW \times PRW, ~CW \times CW\}$. For $P_1, P_2 \in \mathcal{P}$, the reimbursement cost of an item with age $t$ and usage $u$ is modeled by \cite{das2025determination}:
\begin{align}\label{warranty}
   C_{P_1 \times P_2}^{T \times U}(t,u) = \frac{1}{S}\, C_{P_1}^T(t)\, C_{P_2}^U(u).
\end{align}

For the sake of conciseness, here we are presenting the per unit warranty cost under the scanrio $CW\times CW$. So the cost of reimbursement of an item under $CW\times CW$ policy:
\begin{align*}
    C_{CW\times CW}^{T\times U}(t,u)=\begin{dcases}S & \text { if } 0\leq t\leq t_{w_1} ~~0\leq u\leq u_{w_1}\\
    S\left(\frac{t_{w_2}-t}{t_{w_2}-t_{w_1}}\right)&\text{ if } t_{w_1}\leq t\leq t_{w_2} ~~0\leq u\leq u_{w_1}\\
    S\left(\frac{u_{w_2}-u}{u_{w_2}-u_{w_1}}\right)&\text{ if } 0\leq t\leq t_{w_1} ~~u_{w_1}\leq u\leq u_{w_2}\\
  S  \left(\frac{t_{w_2}-t}{t_{w_2}-t_{w_1}}\right)\left(\frac{u_{w_2}-u}{u_{w_2}-u_{w_1}}\right)&\text{ if } t_{w_1}\leq t\leq t_{w_2} ~~u_{w_1}\leq u\leq u_{w_2}.
    \end{dcases}
\end{align*}
It is noted that from the scenario \( CW \times CW \), we can derive all the remaining eight policies by considering special cases. Specifically, by taking \( x_{w_1} = x_{w_2} \), \( CW \) reduces to the \( FRW \) policy, and when \( x_{w_1} = 0 \), \( CW \) reduces to the \( PRW \) policy.  Warranty cost can be expressed as the product of the expected number of failed items and the average reimbursement cost per item:
\begin{align*}
W(t,u,t_{w_1},t_{w_2},u_{w_1},u_{w_2})
= [\text{Expected No. of failures within warranty}] 
\times C_{P_1\times P_2}^{T\times U}(t,u).
\end{align*}

\noindent Expanding this, the warranty cost function becomes
\allowdisplaybreaks
\begin{align*}
W(t,u,t_{w_1},t_{w_2},u_{w_1},u_{w_2}) 
&= M F(t_{w_1},u_{w_1}\mid\mathbf{x}) S \,\mathbb{I}_{[0,t_{w_1})\times[0,u_{w_1})}(t,u) \\
&+ M\big[F(t_{w_2},u_{w_1}\mid\mathbf{x})-F(t_{w_1},u_{w_1}\mid\mathbf{x})\big] 
  S \frac{t_{w_2}-t}{t_{w_2}-t_{w_1}} \,
  \mathbb{I}_{[t_{w_1},t_{w_2})\times[0,u_{w_1})}(t,u) \\
&+ M\big[F(t_{w_1},u_{w_2}\mid\mathbf{x})-F(t_{w_1},u_{w_1}\mid\mathbf{x})\big] 
  S \frac{u_{w_2}-u}{u_{w_2}-u_{w_1}} \,
  \mathbb{I}_{[0,t_{w_1})\times[u_{w_1},u_{w_2})}(t,u) \\
&+ M\big[F(t_{w_2},u_{w_2}\mid\mathbf{x})+F(t_{w_1},u_{w_1}\mid\mathbf{x})-F(t_{w_2},u_{w_1}\mid\mathbf{x})-F(t_{w_1},u_{w_2}\mid\mathbf{x})\big] \\
& \qquad \times S \frac{t_{w_2}-t}{t_{w_2}-t_{w_1}} \frac{u_{w_2}-u}{u_{w_2}-u_{w_1}} \,
  \mathbb{I}_{[t_{w_1},t_{w_2})\times[u_{w_1},u_{w_2})}(t,u),
\end{align*}
where $S$ denotes the unit claim cost, $M$ is the market size, and $\mathbb{I}_{[a,b)\times[c,d)}(t,u)$ is an indicator function.

\subsection{Dissatisfaction Cost Function:}

We consider another cost function, which is the manufacturer's indirect cost to the product. This is called the dissatisfaction cost or penalty cost. In the two-dimensional scenario, when the product fails, consumers have certain expectations about the age and mileage. Suppose that the consumer's expected age and mileage of the product are $L_t$ and $L_u$. This means that when a product fails, if the product's age and the usage are greater than $L_t$ and $L_u$, the consumer is satisfied with the product; otherwise, the manufacturer bears a penalty cost in future sales. For a generic dimension $x \in \{t,u\}$ and $X\in\{T,U\}$, the dissatisfaction cost function is defined as
\begin{align*}
    D^{X
    }(x)=S\times\begin{dcases}
        q_1^X & \text{if } 0\leq x\leq x_{w_1},\\
        q_1^X-(q_1^X-q_2^X)\frac{x-x_{w_1}}{x_{w_2}-x_{w_1}} & \text{if }x_{w_1}< x\leq x_{w_2},\\
        q_2^X\frac{L_x-x}{L_x-x_{w_2}} & \text{if }x_{w_2}< x\leq L_x,\\
        0 & \text{if }x>L_x,
    \end{dcases}
\end{align*}
where  at $x_{w_1}$, per unit cost of dissatisfaction is $S\,q^X_{1}$ and, at $x_{w_2}$, per unit cost of dissatisfaction is $S\,q^X_{2}$ is considered with $0<q^X_{2} < q^X_{1}<1$. Therefore, the dissatisfaction cost can be written as
\small\begin{align*}
    D(t,u,t_{1},t_{2},u_{1},u_{2})=\left\{\text{Expected number of failures in }[t_1,t_2]\times[u_1,u_2]\right\}\times \frac{D^{T}(t_1)+ D^{U}(u_1)}{2}\mathbb{I}_{[t_1,t_2]\times[u_1,u_2]}(t,u)
\end{align*}
\normalsize
\textbf{Case-I: } Items fails in the region $[0, t_{w_1}]\times[0, u_{w_1}]$
\begin{align*}
    D_1(t,u,t_{w_1},t_{w_2},u_{w_1},u_{w_2})=MF(t_{w_1},u_{w_1}\ | \ \mathbf{x})\,\frac{S}{2}\,(q^T_1+q^U_1)\mathbb{I}_{[0,t_{w_1}]\times[0,u_{w_1}]}(t,u).
\end{align*}
\textbf{Case II: } Items fails in the region $[0, t_{w_1}]\times[u_{w_1}, u_{w_2}]$. 
The dissatisfaction cost is defined as
\begin{align*}
    D_2(t,u,t_{w_1},t_{w_2},u_{w_1},u_{w_2})=&M\left[F(t_{w_1},u_{w_2}\ | \ \mathbf{x})-F(t_{w_1},u_{w_1}\ | \ \mathbf{x})\right]\,\frac{S}{2}\,\left[(q^T_1+q^U_1)-(q^U_1-q^U_2)\frac{u-u_{w_1}}{u_{w_2}-u_{w_1}}\right]\\
   & \times\mathbb{I}_{[0,t_{w_1}]\times[u_{w_1},u_{w_2}]}(t,u).
\end{align*} 
\textbf{Case III: } Items fails in the region $[t_{w_1}, t_{w_2}]\times[0, u_{w_1}]$. The dissatisfaction cost in Case III is defined as
\begin{align*}
    D_3(t,u,t_{w_1},t_{w_2},u_{w_1},u_{w_2})=&M\left[F(t_{w_2},u_{w_1}\ | \ \mathbf{x})-F(t_{w_1},u_{w_1}\ | \ \mathbf{x})\right]\,\frac{S}{2}\,\left[(q^T_1+q^U_1)-(q^T_1-q^T_2)\frac{t-t_{w_1}}{t_{w_2}-t_{w_1}}\right]\\
   & \times\mathbb{I}_{[t_{w_1},t_{w_2}]\times[0,u_{w_1}]}(t,u).
\end{align*}
 \textbf{Case IV: } Items fails in the region $[t_{w_1}, t_{w_2}]\times[u_{w_1}, u_{w_2}]$. The dissatisfaction cost is defined as
\small\begin{align*}
    D_4(t,u,t_{w_1},t_{w_2},u_{w_1},u_{w_2})=&M\left[F(t_{w_2},u_{w_2}\ | \ \mathbf{x})+F(t_{w_1},u_{w_1}\ | \ \mathbf{x})-F(t_{w_2},u_{w_1}\ | \ \mathbf{x})-F(t_{w_1},u_{w_2}\ | \ \mathbf{x})\right]\,\frac{S}{2}\,\\
    &\times\left[q^T_1-(q_1^T-q_2^T)\frac{L_t-t}{L_t-t_{w_2}}+q^U_1-(q_1^U-q_2^U)\frac{L_u-u}{L_u-u_{w_2}}\right]\mathbb{I}_{[t_{w_1},t_{w_2}]\times[u_{w_1},u_{w_2}]}(t,u).
\end{align*}
\normalsize\textbf{Case V: } Items fails in the region $[0, t_{w_1}]\times[u_{w_2}, L_u]$. The dissatisfaction cost in  is defined as
\begin{align*}
    D_5(t,u,t_{w_1},t_{w_2},u_{w_1},u_{w_2})=&M\left[F(t_{w_1},L_u\ | \ \mathbf{x})-F(t_{w_1},u_{w_2}\ | \ \mathbf{x})\right]\,\frac{S}{2}\,\left[q^T_1+q^U_2\frac{L_u-u}{L_u-t_{w_2}}\right]\\
   & \times\mathbb{I}_{[0,t_{w_1}]\times[u_{w_2},L_u]}(t,u).
\end{align*}
  \textbf{Case VI: } Items fails in the region $[ t_{w_2},L_t]\times[0, u_{w_1}]$. The dissatisfaction cost in is defined as
\begin{align*}
    D_6(t,u,t_{w_1},t_{w_2},u_{w_1},u_{w_2})=&M\left[F(L_t,u_{w_1}\ | \ \mathbf{x})-F(t_{w_2},u_{w_1}\ | \ \mathbf{x})\right]\,\frac{S}{2}\,\left[q_2^T\frac{L_t-t}{L_t-t_{w_2}}+q^U_1\right]\\
   & \times\mathbb{I}_{[t_{w_2},L_t]\times[0,u_{w_1}]}(t,u).
\end{align*}
 \textbf{Case VII: } Items fails in the region $[t_{w_1}, t_{w_2}]\times[u_{w_2}, L_u]$. The dissatisfaction cost  is defined as
\begin{align*}
    D_7(t,u,t_{w_1},t_{w_2},u_{w_1},u_{w_2})=&M\left[F(t_{w_2},L_u\ | \ \mathbf{x})+F(t_{w_1},u_{w_2}\ | \ \mathbf{x})-F(t_{w_1},L_u\ | \ \mathbf{x})-F(t_{w_2},u_{w_2}\ | \ \mathbf{x})\right]\\
    &\times\frac{S}{2}\left[q_1^T-(q^T_1-q^T_2)\frac{t-t_{w_1}}{t_{w_2}-t_{w_1}}+q_2^U\frac{L_u-u}{L_u-u_{w_2}}\right]\mathbb{I}_{[t_{w_1},t_{w_2}]\times[u_{w_2},L_u]}(t,u).
\end{align*}
 \textbf{Case VIII: } Items fails in the region $[ t_{w_2},L_t]\times[u_{w_1}, u_{w_2}]$. The dissatisfaction cost is defined as
\begin{align*}
    D_8(t,u,t_{w_1},t_{w_2},u_{w_1},u_{w_2})=&M\left[F(L_t,u_{w_2}\ | \ \mathbf{x})+F(t_{w_2},u_{w_1}\ | \ \mathbf{x})-F(t_{w_2},u_{w_2}\ | \ \mathbf{x})-F(L_t,u_{w_1}\ | \ \mathbf{x})\right]\\
    &\times\frac{S}{2}\left[q_2^T\frac{L_t-t}{L_t-t_{w_2}}+q_1^U-(q^U_1-q^U_2)\frac{u-u_{w_1}}{u_{w_2}-u_{w_1}}\right]\mathbb{I}_{[t_{w_2},L_t]\times[u_{w_1},u_{w_2}]}(t,u).
\end{align*}
\textbf{Case IX: } Items fails in the region $[t_{w_2},L_t]\times[u_{w_1},L_u]$
The dissatisfaction cost is defined as
\begin{align*}
    D_9(t,u,t_{w_1},t_{w_2},u_{w_1},u_{w_2})=&M\left[F(L_t,L_u\ | \ \mathbf{x})+F(t_{w_2},u_{w_2}\ | \ \mathbf{x})-F(t_{w_2},L_u\ | \ \mathbf{x})-F(L_t,u_{w_2}\ | \ \mathbf{x})\right]\\
    &\times\frac{S}{2}\left[q_2^T\frac{L_t-t}{L_t-t_{w_2}}+q_2^U\frac{L_u-u}{u_{w_2}-u_{w_1}}\right]\mathbb{I}_{[t_{w_2},L_t]\times[u_{w_2},L_u]}(t,u).
\end{align*}

\noindent The total dissatisfaction cost is constructed by taking the sum of the dissatisfaction costs from all nine possible scenarios.
\begin{align*}
    D(t,u,t_{w_1},t_{w_2},u_{w_1},u_{w_2})=\sum_{i=1}^9D_i(t,u,t_{w_1},t_{w_2},u_{w_1},u_{w_2})
\end{align*}
\begin{figure}[hbt!]
    \centering
    \includegraphics[width=0.8\linewidth]{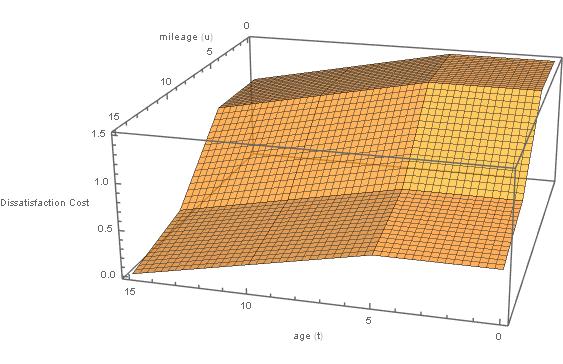}
    \caption{Dissatisfaction cost with age t and mileage u}
    \label{fig:enter-label}
\end{figure}
\section{Optimal Warranty Region}\label{optimal}
The utility function defined in this paper is composed of three non linear functions: the economic benefit function, the warranty cost function and the dissatisfaction cost function. The resulting utility function is then expressed as
\small\begin{eqnarray}
    \mathcal{U}(t, u, t_{w_1},t_{w_2},u_{w_1},u_{w_2}) =  {EB}(t_{w_1},t_{w_2},u_{w_1},u_{w_2}) -{W}(t, u, t_{w_1},t_{w_2},u_{w_1},u_{w_2}) -  {D}(t,u,t_{w_1},t_{w_2},u_{w_1},u_{w_2}). \label{utility}
\end{eqnarray}
\normalsize
By taking the expectation on both sides of (\ref{utility}), we can derive the expected utility function, which will determine the optimal warranty region. The expression is given below
\small\begin{align}
    E[\mathcal{U}(t, u, t_{w_1}, t_{w_2}, u_{w_1}, u_{w_2})] & = EB(t_{w_1},t_{w_2},u_{w_1},u_{w_2}) -E[W(t, u, t_{w_1},t_{w_2},u_{w_1},u_{w_2})] - E[D(t,u,t_{w_1},t_{w_2},u_{w_1},u_{w_2})] \nonumber \\    
    &= EB(t_{w_1},t_{w_2},u_{w_1},u_{w_2}) - \int_{0}^\infty\int_{0}^\infty W(t, u, t_{w_1}, t_{w_2}, u_{w_1}, u_{w_2})~ f(t, u \mid \mathbf{x})~ dt~ du \nonumber\\
    &-\int_{0}^\infty \int_{0}^\infty D(t,u,t_{w_1},t_{w_2},u_{w_1},u_{w_2})~ f(t, u \mid \mathbf{x})~ dt~ du.\label{expected_utility}
\end{align}

\noindent The expected warranty cost derived is given below
\small\begin{align*}
E[W(t, u, t_{w_1},t_{w_2},u_{w_1},u_{w_2})] = &~~~~ M F(t_{w_1},u_{w_1}\ | \ \mathbf{x}) S \times W^{(1)} 
+ M  \big[F(t_{w_2},u_{w_1}\ | \ \mathbf{x}) - F(t_{w_1},u_{w_1}\ | \ \mathbf{x})\big] ~S \times  W^{(2)} \\
& + M \big[F(t_{w_1},u_{w_2}\ | \ \mathbf{x}) - F(t_{w_1},u_{w_1}\ | \ \mathbf{x})\big] ~S \times W^{(3)} + M  \big[F(t_{w_2},u_{w_2}\ | \ \mathbf{x})
 \\
&+ F(t_{w_1},u_{w_1}\ | \ \mathbf{x}) - F(t_{w_2},u_{w_1}\ | \ \mathbf{x}) - F(t_{w_1},u_{w_2}\ | \ \mathbf{x})\big] ~S \times W^{(4)},
\end{align*}
 and 
\begin{align*}
E[D(t, u, t_{w_1},t_{w_2},u_{w_1},u_{w_2})] = &~~~~ M F(t_{w_1},u_{w_1}\ | \ \mathbf{x}) \frac{S}{2} \times D^{(1)}
+ M  \big[F(t_{w_2},u_{w_1}\ | \ \mathbf{x}) - F(t_{w_1},u_{w_1}\ | \ \mathbf{x})\big] ~\frac{S}{2} \times  D^{(2)} \\ &+ M \big[F(t_{w_1},u_{w_2}\ | \ \mathbf{x}) - F(t_{w_1},u_{w_1}\ | \ \mathbf{x})\big] ~\frac{S}{2} \times D^{(3)}+ M  \big[F(t_{w_2},u_{w_2}\ | \ \mathbf{x})  \\ &+ F(t_{w_1},u_{w_1}\ | \ \mathbf{x}) - F(t_{w_2},u_{w_1}\ | \ \mathbf{x}) - F(t_{w_1},u_{w_2}\ | \ \mathbf{x})\big] ~\frac{S}{2} \times D^{(4)}\\
&+ M \left[F(t_{w_1},L_u\ | \ \mathbf{x})-F(t_{w_1},u_{w_2}\ | \ \mathbf{x})\right] \frac{S}{2} \times D^{(5)}\\
&+M\left[F(L_t,u_{w_1}\ | \ \mathbf{x})-F(t_{w_2},u_{w_1}\ | \ \mathbf{x})\right]  \frac{S}{2} \times D^{(6)}\\
&+M \left[F(t_{w_2},L_u\ | \ \mathbf{x})+F(t_{w_1},u_{w_2}\ | \ \mathbf{x})-F(t_{w_1},L_u\ | \ \mathbf{x})-F(t_{w_2},u_{w_2}\ | \ \mathbf{x})\right] \frac{S}{2} \times D^{(7)}\\
&+M \left[F(L_t,u_{w_2}\ | \ \mathbf{x})+F(t_{w_2},u_{w_1}\ | \ \mathbf{x})-F(t_{w_2},u_{w_2}\ | \ \mathbf{x})-F(L_t,u_{w_1}\ | \ \mathbf{x})\right]  \frac{S}{2} \times D^{(8)}\\
&+M \left[F(L_t,L_u\ | \ \mathbf{x})+F(t_{w_2},u_{w_2}\ | \ \mathbf{x})-F(t_{w_2},L_u\ | \ \mathbf{x})-F(L_t,u_{w_2}\ | \ \mathbf{x})\right]  \frac{S}{2} \times D^{(9)},\\
\end{align*}
\normalsize where 
\begin{eqnarray*}
    W^{(1)} &=& \int_{0}^{t_{w_1}} \int_{0}^{u_{w_1}} f(t,u\ | \ \mathbf{x}) \, du \, dt = F(t_{w_1}, u_{w_1}\ | \ \mathbf{x}), \\
   W^{(2)} & = & \int_{t_{w_1}}^{t_{w_2}} \int_{0}^{u_{w_1}} \frac{t_{w_2} - t}{t_{w_2} - t_{w_1}} f(t,u\ | \ \mathbf{x}) \, du \, dt, \\
   W^{(3)} & = & \int_{0}^{t_{w_1}} \int_{u_{w_1}}^{u_{w_2}} \frac{u_{w_2} - u}{u_{w_2} - u_{w_1}}  f(t,u\ | \ \mathbf{x}) \, du \, dt, \\
   W^{(4)} & = & \int_{t_{w_1}}^{t_{w_2}} \int_{u_{w_1}}^{u_{w_2}} \frac{(t_{w_2} - t)(u_{w_2} - u)}{(t_{w_2} - t_{w_1})(u_{w_2} - u_{w_1})}  f(t,u\ | \ \mathbf{x}) \, du \, dt,\\
\end{eqnarray*}
and 
\begin{eqnarray*}
    D^{(1)} &=& \int_{0}^{t_{w_1}} \int_{0}^{u_{w_1}}(q^T_1+q^U_1)~  f(t,u\ | \ \mathbf{x}) \, du \, dt =(q^T_1+q^U_1)~ F(t_{w_1}, u_{w_1}\ | \ \mathbf{x}), \\
   D^{(2)} & = & \int_{t_{w_1}}^{t_{w_2}} \int_{0}^{u_{w_1}}\left[(q^T_1+q^U_1)-(q^U_1-q^U_2)\frac{u-u_{w_1}}{u_{w_2}-u_{w_1}}\right]~  f(t,u\ | \ \mathbf{x}) \, du \, dt, \\
   D^{(3)} & = & \int_{0}^{t_{w_1}} \int_{u_{w_1}}^{u_{w_2}} \left[(q^T_1+q^U_1)-(q^T_1-q^T_2)\frac{t-t_{w_1}}{t_{w_2}-t_{w_1}}\right]~  f(t,u\ | \ \mathbf{x}) \, du \, dt, \\
   D^{(4)} & = & \int_{t_{w_1}}^{t_{w_2}} \int_{u_{w_1}}^{u_{w_2}} \left[q^T_1-(q_1^T-q_2^T)\frac{L_t-t}{L_t-t_{w_2}}+q^U_1-(q_1^U-q_2^U)\frac{L_u-u}{L_u-u_{w_2}}\right]~  f(t,u\ | \ \mathbf{x}) \, du \, dt,\\
   D^{(5)}&=&\int_{0}^{t_{w_1}}\int_{u_{w_2}}^{L_u}\left[q^T_1+q^U_2\frac{L_u-u}{L_u-t_{w_2}}\right]~ f(t,u\ | \ \mathbf{x}) \, du \, dt,\\
   D^{(6)}&=&\int_{t_{w_2}}^{L_t}\int_{0}^{u_{w_1}}\left[q_2^T\frac{L_t-t}{L_t-t_{w_2}}+q^U_1\right]~ f(t,u\ | \ \mathbf{x}) \, du \, dt,\\
   D^{(7)}&=&\int_{t_{w_1}}^{t_{w_2}}\int_{u_{w_2}}^{L_u}\left[q_1^T-(q^T_1-q^T_2)\frac{t-t_{w_1}}{t_{w_2}-t_{w_1}}+q_2^U\frac{L_u-u}{L_u-u_{w_2}}\right]~ f(t,u\ | \ \mathbf{x}) \, du \, dt,\\
   D^{(8)}&=&\int_{t_{w_2}}^{L_t}\int_{u_{w_1}}^{u_{w_2}}\left[q_2^T\frac{L_t-t}{L_t-t_{w_2}}+q_1^U-(q^U_1-q^U_2)\frac{u-u_{w_1}}{u_{w_2}-u_{w_1}}\right]~ f(t,u\ | \ \mathbf{x}) \, du \, dt,\\
   D^{(9)}&=&\int_{t_{w_2}}^{L_t}\int_{u_{w_2}}^{L_u}\left[q_2^T\frac{L_t-t}{L_t-t_{w_2}}+q_2^U\frac{L_u-u}{u_{w_2}-u_{w_1}}\right]~ f(t,u\ | \ \mathbf{x}) \, du \, dt.\\
\end{eqnarray*}








\begin{algorithm}
\caption{Metropolis-Hastings Sampling for $(\alpha, \lambda)$}
\label{algo}
\begin{algorithmic}[1]
\State \textbf{Initialize:} Set $\boldsymbol{\psi} \gets \boldsymbol{\psi}_0$
\For{$i = 1$ to $N$}
    \State Set $\boldsymbol{\psi} \gets \boldsymbol{\psi}_{i-1}$
    \State Sample $\boldsymbol{\delta} \sim \mathcal{N}_2\left( \left(\ln \alpha_T, \ln \lambda_T\ln\alpha_U,\ln\lambda_U, \ln\left[\frac{\theta}{1-\theta}\right]\right), I_{\mathbf{x}^d}^{-1}(\boldsymbol{\psi}) \right)$
    \State Set $\boldsymbol{\psi}^* \gets \left( \exp(\delta_1), \exp(\delta_2),\exp(\delta_3),\exp(\delta_4),\frac{\exp(\delta_5)}{1+\exp(\delta_5)}\right) $
    \State Compute acceptance probability $a^*$:
    \State \quad $a^* = \min\left(1, \dfrac{ \pi(\boldsymbol{\psi}^* \mid \mathbf{x}^d) \cdot \alpha_T^* \cdot \lambda_T^*\cdot\alpha_U^*\cdot\lambda_U^*\cdot(1+\theta^*)^2 }{ \pi(\boldsymbol{\psi} \mid \mathbf{x}^d) \cdot \alpha_T \cdot \lambda_T\cdot\alpha_U\cdot\lambda_U\cdot(1+\theta)^2  } \right)$
    \State With probability $a^*$, set $\boldsymbol{\psi}_i \gets \boldsymbol{\psi}^*$;
    \State otherwise, set $\boldsymbol{\psi}_i \gets \boldsymbol{\psi}$
\EndFor
\end{algorithmic}
\end{algorithm}

\vspace{1em}
After running Algorithm \ref{algo}, we obtain $N$ samples of $\boldsymbol{\psi}$. Discard the first $N_0$ samples as burn-in, and let $k = N - N_0$ be the remaining samples. Using these samples, the posterior predictive PDF and CDF can be written as
\begin{align*}
f(t,u\mid\mathbf{x})=\frac{1}{k}~f(t,u\mid\boldsymbol{\psi}_i),
\end{align*}
and 
\begin{align*}
F(t,u\mid\mathbf{x})=\frac{1}{k}~F(t,u\mid\boldsymbol{\psi}_i).
\end{align*}
The optimal warranty region $(t_{w_1}^*, t_{w_2}^*, u_{w_1}^*, u_{w_2}^*)$ can be obtained by maximizing the expected value of the utility function given in  (\ref{expected_utility}). This is a nonlinear optimization problem in four real continuous variables, which can be solved using the 'optim' function in the R-programming.

\section{Real Life Data and Numerical Example}\label{numerical}

In this section we determine the optimal warranty region using two real data sets. The first data set consists of maintenance records for $n=40$ locomotive traction motors, originally reported in \citet{eliashberg1997calculating}. The second data set is a warranty-claims dataset for starter motors compiled from field failure records. The manufacturer's identity is withheld for confidentiality.

\subsection{Data set 1}

We illustrate the proposed methodology using bivariate failure data obtained from maintenance records of $n=40$ locomotive traction motors. Although the original data are field failures, we adapt them to a right-censored framework to reflect a realistic warranty scenario. Specifically, we assume the life test is terminated when the product age reaches $T_0=5$ years or the usage (mileage) reaches $U_0=2$ (scaled units). Any unit whose age or usage exceeds these thresholds without failure is treated as a right-censored observation. This censoring scheme allows us to analyze the data under warranty-like conditions and incorporate both observed failures and censored lifetimes into the reliability model.
The bivariate observations (age, mileage) are given in Table~\ref{data1}.

\begin{table}[hbt!]\centering
\caption{Bivariate failure data (locomotive traction motors).}
\label{data1}
\begin{tabular}{cccccccccccc}
\hline
No. & Age & Mileage & No. & Age & Mileage & No. & Age & Mileage & No. & Age & Mileage \\
\hline
1 & 1.66 & 0.9766 &10 & 3.35 & 1.3827 & 19 & 1.28 & 0.5922 &28 & 0.01 & 0.0028 \\
2 & 0.35 & 0.2041 &11 & 1.64 & 0.5992 & 20 & 0.31 & 0.1974 &29 & 0.27 & 0.0095  \\
3 & 2.49 & 1.2392 &12 & 1.45 & 0.6925 & 21 & 0.65 & 0.2030 &30 & 2.95 & 1.2600  \\
4 & 1.90 & 0.9889 &13 & 1.70 & 0.7078 & 22 & 2.21 & 1.2532 &31 & 1.40 & 0.8067  \\
5 & 0.27 & 0.0974 &14 & 1.40 & 0.7553 & 23 & 3.16 & 1.4796 &32 & 0.48 & 0.3099 \\
6 & 0.41 & 0.1594 &15 & 1.66 & 0.9766 & 24 & 0.22 & 0.0979 &33 & 0.02 & 0.0105 \\
7 & 0.59 & 0.2128 &16 & 0.29 & 0.0447 & 25 & 2.61 & 1.5062 &34 & 2.09 & 1.2302\\
8 & 0.75 & 0.2158 &17 & 3.40 & 1.6494 & 26 & 0.32 & 0.2062 &35 & --- & --- \\
9 & 2.23 & 1.1187 &18 & 1.60 & 0.7162 & 27 & 3.97 & 1.6888 &36 & --- & --- \\
\hline
\end{tabular}
\end{table}

The maximum likelihood estimates (MLEs) of the model parameters were obtained by maximizing the log-likelihood in (\ref{loglike}). The estimated parameter values (with standard errors) are
$\widehat{\lambda}_T = 1.522,
\widehat{\lambda}_U = 0.722,
\widehat{\eta}_T = 1.015,
\widehat{\eta}_U = 0.930,
\widehat{\theta} = 0.172,$
with corresponding standard errors
$\mathrm{SE}(\widehat{\lambda}_T,\widehat{\lambda}_U,\widehat{\eta}_T,\widehat{\eta}_U,\widehat{\theta})
= (0.2621,\; 0.1356,\; 0.1153,\; 0.1063,\; 0.0224).$

\subsubsection{Optimal warranty region}

\begin{figure}[hbt!]
    \centering
    \includegraphics[width=0.32\linewidth]{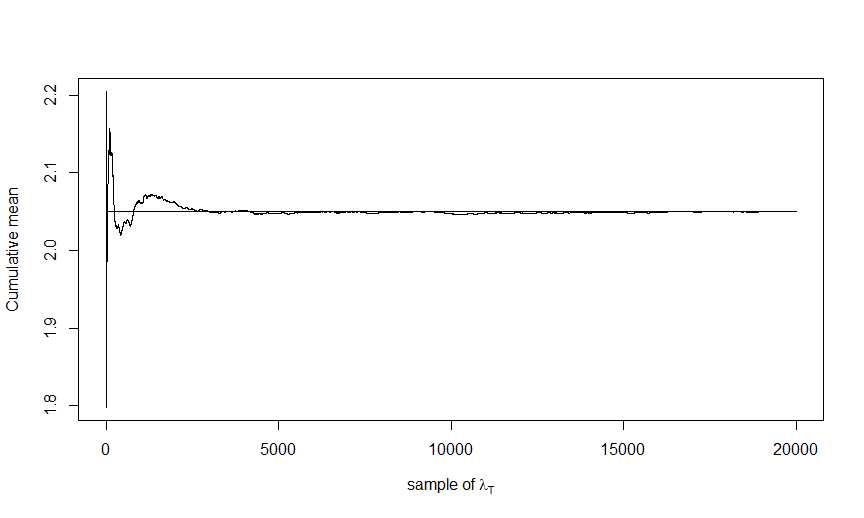}
    \includegraphics[width=0.32\linewidth]{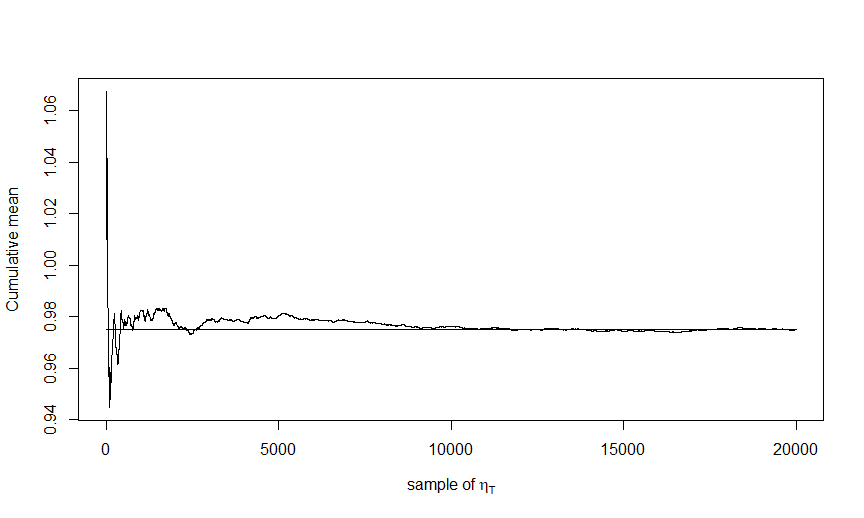}
    \includegraphics[width=0.32\linewidth]{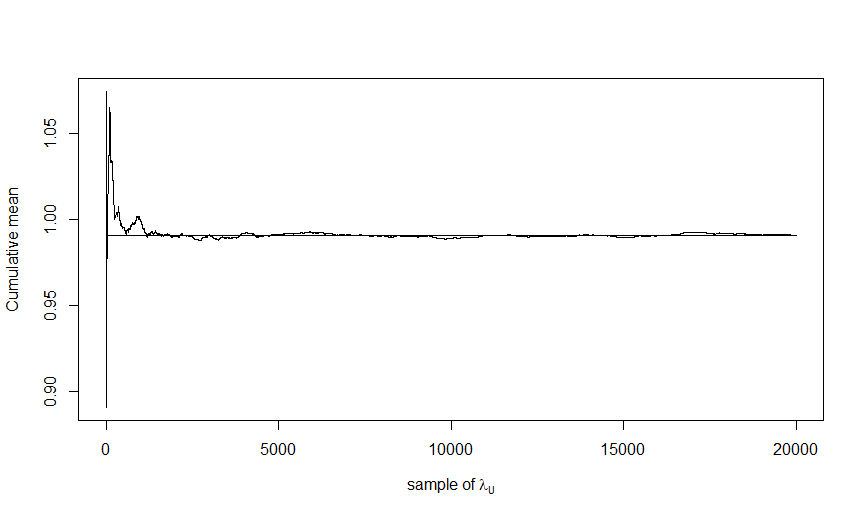}
    \includegraphics[width=0.32\linewidth]{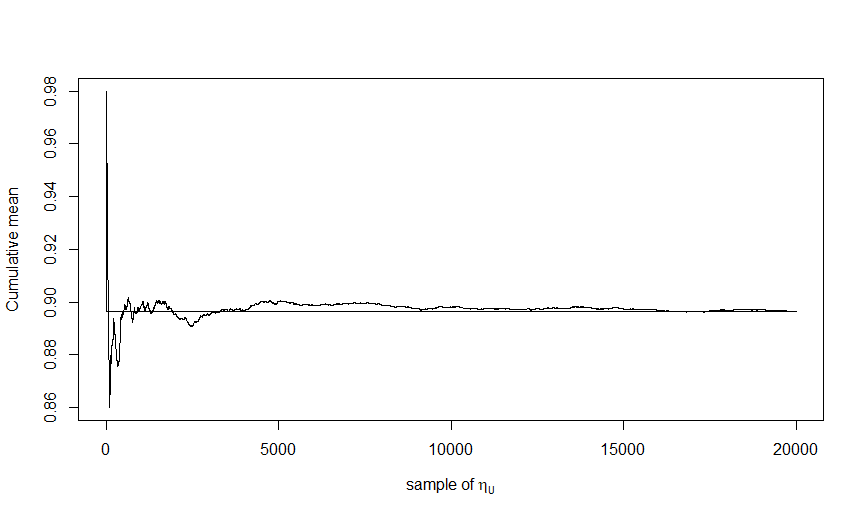}
    \includegraphics[width=0.32\linewidth]{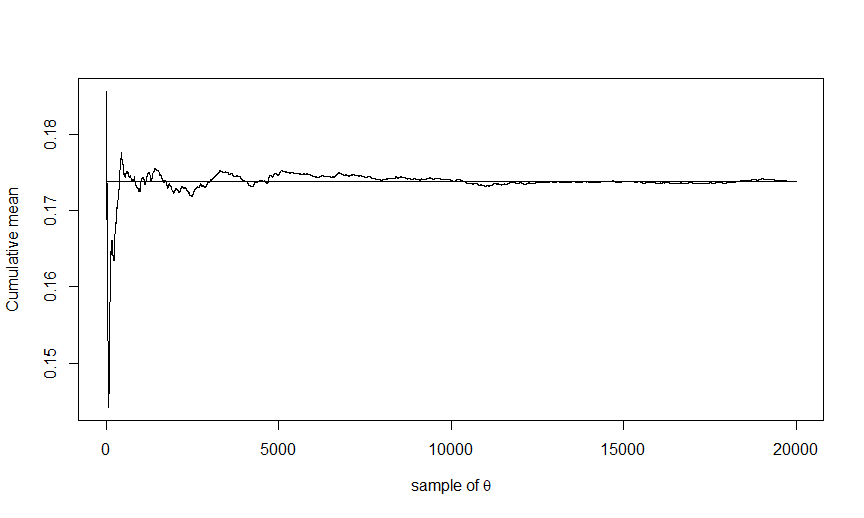}
    \caption{Trace plots for the model parameters (locomotive dataset). The chains exhibit good mixing, supporting reliable posterior inference.}
    \label{trace}
\end{figure}

For the numerical example, we set prior hyperparameters by matching prior means to the MLEs and prior variances to the observed variances via the method of moments. The resulting Gamma and Beta hyperparameters are $a_1=33.719$, $b_1=22.154$, $a_2=28.331$, $b_2=39.239$, $a_3=77.461$, $b_3=76.136$, $a_4=76.540$, $b_4=82.301$, $a_5=48.819$ and $b_5=235.012$. The consumer expected lifetimes as the medians of the posterior predictive marginals are taken as
$L_t = 1.020, L_u = 0.6547$. The FRW reference thresholds (the standard warranty under FRW) are taken as the 0.1 quantiles of the marginal predictive distributions as
$t_w = 0.2006, u_w = 0.0787.$ We calibrate the economic benefit parameters \(A_2\) and \(A_3\) by solving the nonlinear equations
$h(A_2,t_w)=q^*, h(A_3,u_w)=q^*,$
with \(h(\cdot,\cdot)\) defined in (\ref{ratio}) and the chosen reference proportion \(q^*=0.75\). The unique solutions are obtained as 
$A_2 = 10.95, A_3 = 27.91.$ We set the market and cost parameters as follows: unit sale price \(S=700\), production cost \(C=500\), hence per-unit profit \(A_1 = S-C = 200\); market size \(M\) is set as reported in the example (same as earlier). The dissatisfaction proportions are set to
$q_1^T = q_1^U = 0.10, \,q_2^T = q_2^U = 0.05.$
Posterior inference for the lifetime model was obtained by MCMC sampling (Metropolis-Hastings), convergence is assessed via trace plots (Figure~\ref{trace}), which indicate satisfactory stable posterior summaries.
The optimal warranty region was obtained by numerically maximizing the expected utility in (\ref{utility}) over the feasible threshold space \(0<t_{w_1}<t_{w_2}\) and \(0<u_{w_1}<u_{w_2}\). The baseline optimal design and its utility are reported in Table~\ref{cost700}.

\begin{table}[hbt!]
    \centering
    \renewcommand{\arraystretch}{1.1}
    \caption{Two-dimensional optimal warranty design (baseline case, locomotive dataset).}
    \label{cost700}
    \begin{tabular}{ccccS[table-format=6.4]}
    \toprule
     {$t^*_{w_1}$} & {$t^*_{w_2}$} & {$u^*_{w_1}$} & {$u^*_{w_2}$} & {Utility} \\
    \midrule
    0.1435 & 0.9373 & 0.1105 & 0.2048 & 188.5351 \\
    \bottomrule
     \end{tabular}
\end{table}

\subsubsection{Sensitivity analyses}
In this Section, we study the effect of $S$, $q_1$ and $q_2$ in the optimal solution of the warranty region. The other values of the components which are not mentioned in the tables, are kept fixed.
\paragraph{Effect of selling price \(S\).}
Table~\ref{S} reports the optimal warranty design for several values of the sale price \(S\). As \(S\) increases, the optimal warranty region tends to shrink in both age and usage dimensions, while the utility decreases monotonically. This is consistent with the intuition that higher prices alter the trade-off between offering attractive warranties and incurring expected warranty and dissatisfaction costs.

\begin{table}[hbt!]
    \centering
    \renewcommand{\arraystretch}{1.1}
    \caption{Optimal warranty design for varying sale price \(S\) (locomotive dataset).}
    \label{S}
    \begin{tabular}{lccccS[table-format=6.4]}
    \toprule
    $S$ & {$t^*_{w_1}$} & {$t^*_{w_2}$} & {$u^*_{w_1}$} & {$u^*_{w_2}$} & {Utility} \\
    \midrule
    300 & 0.1698 & 1.0101 & 0.1276 & 0.2387 & 193.9477  \\
    400 & 0.1591 & 0.9597 & 0.1226 & 0.2314 & 192.4521  \\
    500 & 0.1504 & 0.9578 & 0.1169 & 0.2187 & 191.0800  \\
    600 & 0.1519 & 0.9477 & 0.1117 & 0.2137 & 189.7709  \\
    800 & 0.1398 & 0.9219 & 0.1065 & 0.1976 & 187.3635  \\
    900 & 0.1392 & 0.9199 & 0.1047 & 0.1938 & 186.1933  \\
   1000 & 0.1368 & 0.9072 & 0.1022 & 0.1893 & 185.0997  \\
   1100 & 0.1311 & 0.8836 & 0.0832 & 0.1832 & 184.0271  \\
    \bottomrule
     \end{tabular}
\end{table}

\paragraph{Effect of benefit-calibration proportions \(q_1\) and \(q_2\).}
Table~\ref{q} reports the optimal warranty thresholds for a grid of \((q_1,q_2,
)\) combinations.  Increasing either \(q_1\) or \(q_2\), it is seen that the value of utility also increases.

\begin{table}[hbt!]
    \centering
    \renewcommand{\arraystretch}{1.1}
    \caption{Optimal warranty design for different values of \(q_1\) and \(q_2\) (locomotive dataset).}
    \label{q}
    \begin{tabular}{ccccccccS[table-format=6.4]}
    \toprule
  $q_1$  & $q_2$ & $A_2$ & $A_3$ & {$t^*_{w_1}$} & {$t^*_{w_2}$} & {$u^*_{w_1}$} & {$u^*_{w_2}$} & {Utility} \\
    \midrule
    0.75 & 0.6  & 10.9524 & 10.3021 & 0.2674 & 0.4968 & 0.0621 & 1.3975 & 184.5516 \\
    0.75 & 0.9  & 10.9524 & 55.8273 & 0.2788 & 0.4577 & 0.0552 & 0.0552 & 187.3107 \\
    0.9  & 0.6  & 21.9048 & 10.3021 & 0.1344 & 0.3092 & 0.3143 & 1.2568 & 189.6405 \\
    0.9  & 0.75 & 21.9048 & 27.9136 & 0.1485 & 0.3298 & 0.0615 & 0.3732 & 192.9109 \\
    0.9  & 0.9  & 21.9048 & 55.8273 & 0.0930 & 0.6750 & 0.0680 & 0.0680 & 194.7369 \\
    \bottomrule
     \end{tabular}
\end{table}

\subsection{Data set 2}

The second dataset contains bivariate failure data for starter motors with variables:
\begin{enumerate}
  \item \textit{Age:} measured in days after installation (scaled by 100 in Table~\ref{data2}), and
  \item \textit{Usage:} measured in operating hours (scaled by 100 in Table~\ref{data2}).
\end{enumerate}
The warranty period is defined as one year or 1000 operating hours, whichever occurs first. The scaled observations are shown in Table~\ref{data2}.

\begin{table}[hbt!]
\centering
\caption{Bivariate failure data (starter motors), scaled by 100.}
\label{data2}
\begin{tabular}{cccccccccccc}
\hline
No. & Age & Usage & No. & Age & Usage & No. & Age & Usage & No. & Age & Usage \\
\hline
1 & 0.01 & 0.02 & 12 & 1.60 & 6.33 & 23 & 2.91 & 6.90 &34 & 2.39 & 5.84 \\
2 & 0.09 & 0.32 & 13 & 1.75 & 4.91 & 24 & 3.15 & 8.62 &35 & 2.45 & 7.55 \\
3 & 0.42 & 1.26 & 14 & 1.85 & 6.38 & 25 & 3.29 & 4.81 &36 & 2.52 & 6.99 \\
4 & 0.63 & 3.20 & 15 & 2.02 & 2.31 & 26 & 0.68 & 2.02 &37 & 2.66 & 7.27 \\
5 & 0.66 & 2.00 & 16 & 2.06 & 5.51 & 27 & 0.80 & 2.37 &38 & 2.68 & 9.40 \\
6 & 0.76 & 1.01 & 17 & 2.41 & 7.42 & 28 & 0.91 & 2.34 &39 & 2.81 & 7.68 \\
7 & 0.81 & 2.90 & 18 & 2.44 & 9.83 & 29 & 1.01 & 3.02 &40 & 2.96 & 8.10 \\
8 & 0.84 & 2.92 & 19 & 2.48 & 7.90 & 30 & 1.33 & 3.58 &41 & 3.02 & 7.86 \\
9 & 1.04 & 3.71 & 20 & 2.70 & 7.13 & 31 & 1.44 & 3.88 &42 & 3.51 & 7.57 \\
10 & 1.25 & 7.00 & 21 & 2.80 & 7.96 & 32 & 1.94 & 5.77 &43 & 3.60 & 6.23 \\
11 & 1.39 & 3.75 & 22 & 2.89 & 8.85 & 33 & 2.02 & 5.69 &    &     &     \\
\hline
\end{tabular}
\end{table}

\subsubsection{Data analysis and model fit}

\begin{figure}[hbt!]
    \centering
    \includegraphics[width=0.4\linewidth]{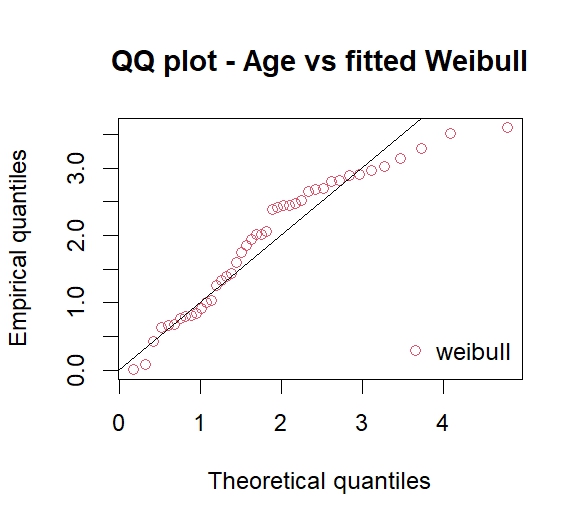}
    \includegraphics[width=0.4\linewidth]{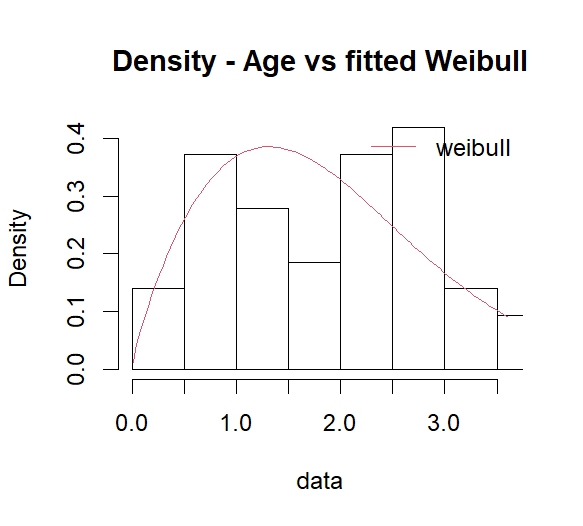} \\
    \includegraphics[width=0.4\linewidth]{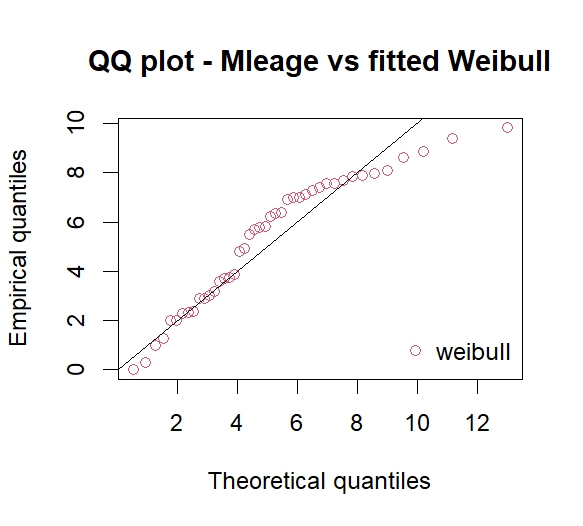}
    \includegraphics[width=0.4\linewidth]{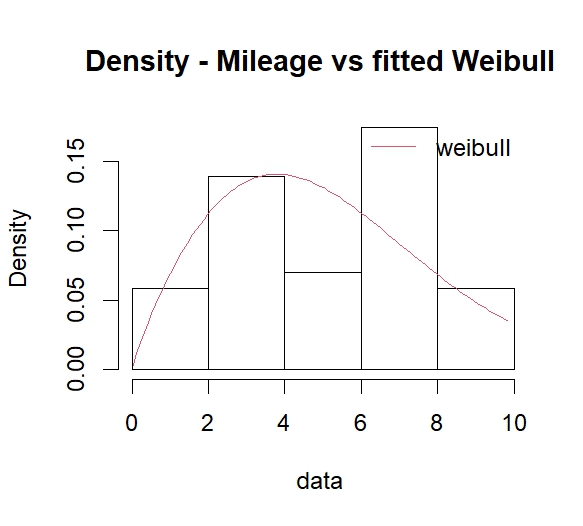}
    \caption{Graphical assessment of Weibull fits for usage (top row) and age (bottom row): fitted densities (left) and QQ plots (right).}
    \label{fig}
\end{figure}

We fitted marginal Weibull distributions to the age and usage variables. The MLEs are
$\widehat{\lambda}_T = 2.079, \widehat{\eta}_T = 1.788;
\widehat{\lambda}_U = 5.797, \widehat{\eta}_U = 1.846.$
Graphical assessments (histograms with fitted densities and QQ plots) are presented in Figure~\ref{fig}; the Anderson--Darling test yields $p$-values $0.2907$ (age) and $0.2226$ (usage), supporting the Weibull marginal fits. The Pearson correlation between age and usage is estimated at $0.8539$, indicating strong positive dependence.
Maximizing the log-likelihood function given in (\ref{loglike}), we get the MLEs
$\hat{\boldsymbol{\psi}} = (2.055,\; 5.869,\; 1.900,\; 1.879,\; 0.282),$
with observed parameter variances
$\widehat{\mathrm{Var}}(\hat{\boldsymbol{\psi}})=(0.0286,\; 0.2412,\; 0.0592,\; 0.0588,\; 0.4199).$
\subsubsection{Optimal-region computation under different censoring}
We set prior hyperparameters by matching prior means to the MLEs and prior variances to the observed variances via the method of moments. The resulting Gamma and Beta hyperparameters are $a_1=147.658$, $b_1=71.853$, $a_2=142.807$, $b_2=24.333$, $a_3=60.980$, $b_3=32.095$, $a_4=60.045$, $b_4=31.956$, $a_5=21.679$ and $b_5=55.197$.
For illustration we again set \(S=700\), \(C=500\) (hence \(A_1=200\)) and \(q^*=0.75\). Solving the calibration equations produces \(A_2=11.41033\) and \(A_3=28.95686\). All other constants (\(M,q^T_i,q^U_i,q_t^*,q_u^*\)) are set as in Data set 1. Table~\ref{censoring} reports the optimal warranty designs for a variety of censoring cutoffs \((T_0,U_0)\) and the corresponding utility values. These results demonstrate how the optimal thresholds change under different censoring schemes.

\begin{table}[hbt!]
    \centering
    \renewcommand{\arraystretch}{1}
    \caption{Optimal warranty design for different censoring setups (starter-motor dataset).}
    \label{censoring}
    \begin{tabular}{ccccccccccS[table-format=6.4]}
    \toprule
  $T_0$  &$U_0$ &$t_w$&$u_w$ &$A_2$& $A_3$&   {$t^*_{w_1}$} & {$t^*_{w_2}$} & {$u^*_{w_1}$} & {$u^*_{w_2}$} & {Utility} \\
    \midrule
    2 & 5 & 0.4873 & 1.2709 & 4.5093 & 1.7289 & 0.5982 & 1.4113 & 1.6029 & 2.4515 & 183.5146\\
    3 & 5 & 0.6516 & 1.4899 & 3.3720 & 1.4747 & 0.8328 & 1.4164 & 1.8771 & 2.8546 & 181.9033 \\
    4 & 5 & 0.4005 & 1.1379 & 5.4857 & 1.9309 & 0.5274 & 0.4693 & 1.4918 & 1.2976 & 186.6935\\
    2 & 7 & 0.5217 & 1.4970 & 4.2112 & 1.4677 & 0.6629 & 1.2812 & 1.9022 & 3.0453 & 183.5189 \\
    3 & 7 & 0.5806 & 1.6791 & 3.7846 & 1.3086 & 0.7154 & 1.6545 & 2.0760 & 3.0303 & 182.2797 \\
    4 & 7 & 0.5781 & 1.6516 & 3.8006 & 1.3303 & 0.7466 & 1.2979 & 2.1214 & 3.5654 & 183.8244 \\
    2 & 10 & 0.6115 & 1.9229 & 3.5928 & 1.1427 & 0.7789 & 1.3505 & 2.3784 & 3.4568 & 180.9426 \\
    3 & 10 & 0.6144 & 1.7772 & 3.5759 & 1.2364 & 0.6834 & 3.1230 & 2.1457 & 2.9392 & 181.2093 \\
    4 & 10 & 0.6009 & 1.6930 & 3.6566 & 1.2978 & 0.5952 & 3.2108 & 2.0786 & 2.8126 & 180.8729 \\
    \bottomrule
     \end{tabular}
\end{table} 
\section{Conclusion}\label{conclusion}
In this work, we propose dissatisfaction cost for a two-dimensional warranty scenario. We determine the optimal warranty region by the Bayesian approach under a two-dimensional combined policy using three cost functions: the economic benefit function, the warranty cost function, and the dissatisfaction cost function. A bivariate ME model is used to model positively correlated failure time data, where age and usage are marginally Weibull distributed. In numerical analysis, two different real-life  data sets are analyzed.

The work can be extended to an $n$-dimensional warranty. If $X_1,\ldots,X_n$ are random variables of $n$ measurable quantities to the failure of the product, the dissatisfaction cost can be written as
\small\begin{align*}
    D_{W_1\times\cdots\times W_n}^{(X_1\times\cdots\times X_n)}(x_1,\ldots,x_n,x_{11},x_{12},\ldots,x_{n1},x_{n2})=&\left\{\text{Expected number of failures in }[x_{11},x_{12}]\times\cdots\times[x_{n1},x_{n2}]\right\}S\\
    & \times\frac{D^{X_1}(x_1)+\cdots+ D^{X_n}(x_n)}{2}\mathbb{I}_{[x_{11},x_{12}]\times\cdots\times[x_{n1},x_{n2}]}(x_1,\ldots,x_n).
\end{align*}
\normalsize
Also, in this work, we use linear pro-rated compensation for each individual scale. However, the work can be extended to use a non-linear function for pro-rated compensation for the individual scales (see \citet{sen2022determination}).


\bibliographystyle{apalike}
\bibliography{BaysWarranty}
\vspace{0.5cm}
\appendix
\noindent\textbf{Appendix}\\
\textbf{Proof of Lemma \ref{lem:score}:}
\begin{proof}
    \begin{align}\label{ul}
        l_i(\boldsymbol{\psi})=\delta_i\log f(t_i,u_i)+(1-\delta_i)\log[1-F(T_0,U_0)]
    \end{align}
    By differentiating (\ref{ul}) with respect to $\psi_u, \, u = 1,2,3,4,5$, we have
    \begin{align}\label{lemma_2_dl}
    \partial_{\psi_u}l_i(\boldsymbol{\psi})=\left[\delta_i \partial_{\psi_u}\log f(t_i,u_i\mid\boldsymbol{\psi})+(1-\delta_i) \partial_{\psi_u}\log[1-F(T_0-U_0\mid\boldsymbol{\psi})]\right]
    \end{align}
    Taking the expectation of (\ref{lemma_2_dl}), we have
    \begin{align*}
    E\left[\frac{\partial l_i(\boldsymbol{\psi})}{\partial \psi_u}\right] =&\int_0^{T_0}\int_0^{U_0}\partial_{\psi_u}\log f(t,u\mid\boldsymbol{\psi})f(t,u\mid\boldsymbol{\psi})~dt~du+[1-F(T_0,U_0\mid\boldsymbol{\psi})]\partial_{\psi_u}\log[1-F(T_0-U_0\mid\boldsymbol{\psi})]\\ =&\int_0^{T_0}\int_0^{U_0}\partial_{\psi_u}f(t,u\mid\boldsymbol{\psi})~dt~du+\partial_{\psi_u}[1-F(T_0,U_0\mid\boldsymbol{\psi})] \\
=\,&n\left[\partial_{\psi_u}F(T_0,U_0\mid\boldsymbol{\psi})-\partial_{\psi_u}[F(T_0,U_0\mid\boldsymbol{\psi})]  \right]\\
=\,&0
    \end{align*}
    This completes the proof.
    \end{proof}\\
    
\noindent \textbf{Proof of Lemma \ref{lemma_2}}
\begin{proof}
    \begin{align}\label{lemma_2_d2l}
       \partial_{\psi_u}l_i(\boldsymbol{\psi})&=\delta_i \partial_{\psi_u}\log f(t_i,u_i\mid\boldsymbol{\psi})+(1-\delta_i) \partial_{\psi_u}\log[1-F(T_0-U_0\mid\boldsymbol{\psi})]\nonumber\\
        &=\frac{\delta_i}{f(t_i,u_i\mid\boldsymbol{\psi})}\partial_{\psi_u}f(t_i,u_i\mid\boldsymbol{\psi})+ \frac{1-\delta_i}{1-F(t_0,U_0\mid\boldsymbol{\psi})}\partial_{\psi_u}[1-F(T_0-U_0\mid\boldsymbol{\psi})]
    \end{align}
    By differentiating (\ref{lemma_2_d2l}) with respect to $\psi_v, \, v = 1,2,3,4,5$, we have
     \begin{align}\label{lemma_2_d2}
       \partial^2_{\psi_u\psi_v}l_i(\boldsymbol{\psi})&=\frac{\delta_i}{f(t_i,u_i\mid\boldsymbol{\psi})}\partial^2_{\psi_u\psi_v}f(t_i,u_i\mid\boldsymbol{\psi})+ \frac{1-\delta_i}{1-F(t_0,U_0\mid\boldsymbol{\psi})}\partial^2_{\psi_u\psi_v}[1-F(T_0-U_0\mid\boldsymbol{\psi})]\nonumber\\
        &- \frac{\delta_i}{[f(t_i,u_i\mid\boldsymbol{\psi})]^2}\partial_{\psi_u}f(t_i,u_i\mid\boldsymbol{\psi}\mid\boldsymbol{\psi})\partial_{\psi_v}f(t_i,u_i\mid\boldsymbol{\psi})\nonumber\\
        &- \frac{(1-\delta_i)}{[1-F(t_0,U_0\mid\boldsymbol{\psi})]^2}\partial_{\psi_u}[1-F(T_0-U_0\mid\boldsymbol{\psi})]\partial_{\psi_v}[1-F(T_0-U_0\mid\boldsymbol{\psi})].
    \end{align}
    Similarly from Lemma \ref{lem:score}, we get
    \begin{align*}
        E\left[ \frac{\delta_i}{f(t_i,u_i\mid\boldsymbol{\psi})}\partial^2_{\psi_u\psi_v}f(t_i,u_i\mid\boldsymbol{\psi})+ \frac{1-\delta_i}{1-F(t_0,U_0\mid\boldsymbol{\psi})}\partial^2_{\psi_u\psi_v}[1-F(T_0-U_0\mid\boldsymbol{\psi})]\right]=0
    \end{align*}
    The 2nd part of (\ref{lemma_2_d2}) can be written as
  \footnotesize  \begin{align*}
       & - \frac{\delta_i}{[f(t_i,u_i\mid\boldsymbol{\psi})]^2}\partial_{\psi_u}f(t_i,u_i\mid\boldsymbol{\psi})\partial_{\psi_v}f(t_i,u_i\mid\boldsymbol{\psi})- \frac{(1-\delta_i)}{[1-F(t_0,U_0\mid\boldsymbol{\psi})]^2}\partial_{\psi_u}[1-F(T_0-U_0\mid\boldsymbol{\psi})]\partial_{\psi_v}[1-F(T_0-U_0\mid\boldsymbol{\psi})]\\
       =& -\delta_i\partial_{\psi_u}\log f(t_i,u_i\mid\boldsymbol{\psi})\partial_{\psi_v}\log f(t_i,u_i\mid\boldsymbol{\psi})+(1-\delta_i) \partial_{\psi_u}\log[1-F(T_0-U_0\mid\boldsymbol{\psi})]\partial_{\psi_v}\log[1-F(T_0-U_0\mid\boldsymbol{\psi}\mid\boldsymbol{\psi})]\\
    \end{align*}
 \normalsize   Now, taking the expectation of (\ref{lemma_2_d2}), we get
    \begin{align*}
         E\left[ \frac{\partial^2 l_i(\boldsymbol{\psi})}{\partial\psi_u\partial\psi_v}\right]=& -\left[\int_0^{T_0}\int_0^{U_0}\partial_{\psi_u}\log f(t,u\mid\boldsymbol{\psi})\partial_{\psi_v}\log f(t,u\mid\boldsymbol{\psi})~f(t,u\mid\boldsymbol{\psi})~dt~du\right.\\
         &\left.+[1-F(T_0,U_0\mid\boldsymbol{\psi})] \partial_{\psi_u}\log[1-F(T_0-U_0\mid\boldsymbol{\psi})]\partial_{\psi_v}\log[1-F(T_0-U_0\mid\boldsymbol{\psi})]\right]
    \end{align*}
        This completes the proof.
\end{proof}
\end{document}